\documentclass[iop,apj]{emulateapj}
\slugcomment{{\sc \underline{Accepted for publication in Geoscience Letters:}} July 7, 2016}
\usepackage{amsmath,amssymb,amstext}
\usepackage[breaklinks,colorlinks,citecolor=blue,linkcolor=magenta]{hyperref} 

\usepackage[all]{hypcap} 

\usepackage{natbib,color,times}
\shorttitle{Gravity Waves during Sudden Warmings}
\shortauthors{Yi\u{g}it and Medvedev}
\graphicspath{{/Users/erdal/Research/GWsswGL/Figures/}}

\begin{document}
\title{\textbf{Role of gravity waves in vertical coupling during sudden stratospheric warmings}}

\author{Erdal Y\.{I}\u{g}\.{I}t\altaffilmark{1}, Alexander S. Medvedev\altaffilmark{2,3}}


\altaffiltext{1}{Department of Physics and Astronomy,
 George Mason University, Fairfax, Virginia, USA.}

\altaffiltext{2}{Max Planck Institute for Solar System Research, 
G\"ottingen, Germany.}

 \altaffiltext{3}{Institute of Astrophysics, Georg-August University,
G\"ottingen, Germany.}

\begin{abstract} 
  Gravity waves are primarily generated in the lower atmosphere, and can reach thermospheric
  heights in the course of their propagation.  This paper reviews the recent progress in
  understanding the role of gravity waves in vertical coupling during sudden
  stratospheric warmings.  Modeling of gravity wave effects is briefly reviewed, and the
  recent developments in the field are presented. Then, the impact of these waves on the
  general circulation of the upper atmosphere is outlined. Finally, the role of gravity
  waves in vertical coupling between the lower and the upper atmosphere is discussed in the
  context of sudden stratospheric warmings.
\end{abstract}


\section{Introduction}

The lower atmosphere, where meteorological processes take place, is the primary source of
internal atmospheric waves: gravity waves (GWs), planetary (Rossby) waves, and solar tides.
These waves can propagate upward and influence the dynamics and thermal state of the middle
and upper atmosphere \citep[see e.g., the reviews of][]{FrittsAlexander03, Lastovicka06,
  YigitMedvedev15}.  Waves transfer their energy and momentum to the
mean flow via breaking and dissipative processes, such as radiative damping, eddy viscosity,
nonlinear diffusion, molecular diffusion and thermal conduction, and ion drag
\citep{Yigit_etal08}.  Sudden stratospheric warmings (SSWs) are spectacular events that
disturb the circulation in the winter hemisphere. They affect not only the stratosphere, but
their influence extends to the mesosphere and thermosphere.  In the upper atmosphere, plasma
processes, such as Joule and auroral heating, ion friction are important processes that
shape the morphology and dynamics. Thus, interactions between the lower and upper atmosphere
should be considered within the framework of the atmosphere--ionosphere system.

Such coupled upper atmosphere--ionosphere system is subject to the following internal
and external influences:
\begin{itemize}
\item Meteorological effects that encompass internal wave
impacts and transient processes of lower atmospheric origin,
\item Internal processes due to nonlinearity,
\item Space weather effects that are associated with the solar and magnetospheric phenomena.
\end{itemize}

Among the meteorological effects, we distinguish a direct influence of internal GWs on the
upper regions of the atmosphere. Although transient events such as SSWs are technically
categorized as stratospheric processes, and, thus take place above the region of
weather-dominated phenomena, they are often referred to as meteorological effects in the
context of the upper atmosphere research.

The thermosphere--ionosphere system is highly nonlinear. In the real atmosphere, ion and
neutral parameters vary simultaneously, and the resulting changes in the heating ought to
contain higher order terms, which is indicative of the nonlinear nature of 
the system \citep{YigitRidley11a}.  The atmosphere-ionosphere system is
subject to the influence of space weather, which can enhance these nonlinear processes and
impact the upper atmosphere \citep[][and references therein]{Prolss11}.

In this paper, we report on the recent advances in understanding the
meteorological effects in the upper atmosphere, focusing 
primarily on the links between SSWs, small-scale GWs, and 
thermosphere--ionosphere dynamics.

\section{Internal gravity waves} 
\label{sec:intern-grav-waves} 

Internal gravity waves are characteristic features of all stably stratified planetary atmospheres.
GWs in the upper atmosphere have been studied for more than 50 years since the
early work of \citet{Hines60}. Their importance for the general circulation of the middle atmosphere
has been greatly appreciated \citep[e.g.,][]{GarciaSolomon85, Becker11}.
However, despite the previous theoretical approaches to GW propagation into the thermosphere
\citep{Klostermeyer72,HickeyCole88}, only since recently, the role
of GWs in coupling the lower and upper atmosphere is being increasingly acknowledged
\citep{Yigit_etal09, VadasLiu09, FrittsLund11, Yigit_etal12b, Miyoshi_etal14,Hickey_etal11,
  Hickey_etal10b, Heale_etal14}.

GWs are always present in the lower and upper atmosphere,
however their amplitudes and dynamical importance differ with height.  Wave energy is
proportional to air density, and, therefore, a conservatively propagating harmonic has a
larger amplitude in regions with lower density. In the troposphere, GW amplitudes are
relatively small, however, their dynamical importance increases with height, and can no
longer be neglected in the middle and upper atmosphere.

We next discuss basic principles of how GW processes are represented in atmospheric models,
reviewing the underlying assumptions and limitations.

\subsection{Principles of parameterization of gravity wave processes in 
global atmosphere models}
\label{sec:princ-grav-wave}

Spatial scales of GWs are considerably smaller than the planetary radius. Their sources are highly
intermittent, and propagation is strongly dispersive. Therefore, the GW field in the thermosphere is
highly irregular and transient. Unlike with distinct larger-scale planetary waves, it appears as an
ever changing ``sea of waves" with occasional well-defined and detectable packets. In many
applications, such chaotic wave field and its influence on the larger-scale flow can be conveniently
described in terms of statistical quantities devoid of the phase information.  Examples of the most
widely used statistical characteristics for the GW field are the variance
$\overline{\phi^{\prime 2}}$, vertical flux of horizontal momentum
$\overline{{\bf u}^\prime w^\prime}$, sensible heat flux $\overline{w^\prime T^\prime}$, etc, where
$w^\prime$, $T^\prime$, and $\phi^\prime$ are the deviations of vertical velocity, temperature and
of any field variable from the corresponding mean values, respectively.

General circulation models (GCMs) have spatial resolutions usually much coarser than the
scales of GWs. Only few GCMs have endeavored to perform simulations with grids small enough
in an attempt to resolve at least a part of the GW spectrum
\citep[e.g.,][]{MiyoshiFujiwara08, Miyoshi_etal14}. In most simulation studies, the effects
of subgrid-scale GWs have to
be parameterized. This means that \\
(1) the average effects must be presented in terms of statistical quantities similar to the
described above, and the quantities have to be functions of the background flow. In other
words, the parameterization has to self-consistently capture responses of the wave field to
the evolution of the resolved
larger-scale flow. \\
(2) Parameterizations should preferably be based on first principles, that is, they
  should rely on rigorous laws of physics rather than on a set of empirically introduced
(tuning) parameters. Obviously, no parameterization can be devoid of such parameters as they
are a substitute for an unknown. But the lesser the number of tunable
  parameters, the more sophisticated the parameterization is. \\
(3) Parameterizations must be verifiable. This condition means that they have to
provide quantities, which can be compared with observations. For instance, GW-induced
heating/cooling rates are hard to measure, but temperature variances
$\overline{T^{\prime 2}}$ can be.

\subsection{Assumptions and limitations in gravity wave parameterizations}
\label{sec:assumptions}

In modeling, it is assumed that the majority of GWs are generated in the lower
atmosphere. Amplitudes of those excited in the upper layers and propagating downward
decrease exponentially with height together with their influence on the mean
flow. Therefore, (1) only harmonics propagating upward are considered in
parameterizations. This assumption allows one to omit a detailed consideration of the wave
reflection, and to (2) apply the Wentzel-Kramers-Brillouin (WKB) approximation. Under the
WKB method, (3) only those harmonics are considered whose vertical wavelengths are much
shorter than vertical variations of the background fields. Mathematically, the latter can be
expressed as $k_zH\gg 1$, where $k_z$ is the vertical wavenumber and $H$ is the
density scale height. This limitation becomes very restrictive in the thermosphere, because
fast (and long vertical-wavelength) harmonics have more chances to penetrate from
tropospheric heights. In the real world, GWs propagate obliquely with respect to the
surface. However, because $k_z\gg k_h$ for most harmonics, $k_h$ being the horizontal
  wavenumber, parameterizations (4) usually assume vertical-only propagation. Limitations
of this approximation in the middle atmosphere have been recently discussed in the work by
\citet{Kalisch_etal14}, and higher-order effects have been found with a scheme employing ray
tracing \citep{Song_etal07}.  A special care should be taken with parameterizations
extending to the thermosphere, where longer vertical wavelength harmonics (lower $k_z$) tend
to propagate to from below. In other words, all gravity waves accounted for by a
parameterization must remain within their grid columns. Finally, (5) all column-based
parameterizations employ a steady state approximation. That is, transient processes of wave
propagation assume an instantaneous response to changes in the
forcing below. This approximation is suitable for modeling the
general circulation, however, implications of time delay due to the finite group speed of
wave packets should be carefully weighted for simulations of more rapid processes.

Parameterizations compute vertical profiles of a specified statistical quantity characterizing the
GW field, such as horizontal velocity variance $\overline{u^{\prime 2}}$
\citep[e.g.,][]{MedvedevKlaassen95}, or vertical flux of horizontal momentum
$\overline{u^\prime w^\prime}$ \citep[e.g.,][]{Yigit_etal08}. The former is convenient for
comparison with observations of GW spectra. The latter is physically more lucid, because
$\rho \overline{u^\prime w^\prime}$ is an invariant in a non-dissipative atmosphere. In GCMs,
sources are specified by (1) prescribing the corresponding quantity at a certain level $z_s$ in the
lower atmosphere, or (2) calculating it interactively using large-scale fields resolved by the model
as an input. The latter is sometimes called ``parameterization of gravity wave sources".  Because
mechanisms of wave excitation in the lower atmosphere are numerous, each requires a separate
approach. To date, physically based schemes suitable for GCMs have been developed for
GWs excited by convection \citep{ChunBaik02, Beres_etal04}, flow over topography
  \citep{McFarlane87}, and fronts \citep{CharronManzini02}. In most other modeling studies,
spectra at a source level are prescribed based on observational constraints, or simply tuned to
obtain desired simulated fields. A comprehensive comparison of GW fluxes in observations and
modeling has recently been performed by \citet{Geller_etal13}.  Although many GCMs use
  time-independent source spectra, GW excitation can undergo large changes during transient events,
  such as SSWs. Therefore, the importance of such variations should be explored and
  their possible impacts on the general circulation have to be taken into account in whole
  atmosphere GCMs.

In the middle atmosphere, the main mechanism of GW obliteration is nonlinear breaking and/or
saturation that occurs when amplitudes become large. Therefore, most  GW parameterizations
developed for middle atmosphere GCMs (starting from that of \citet{Lindzen81}) have in common that
they terminate harmonics, whose amplitudes reach a certain instability
threshold. Exceptions are the approaches of \citet{Hines97a} (``Doppler spread") and
\citet{MedvedevKlaassen95} (``nonlinear diffusion"), which sought to describe the underlying
physics. The former is based on the assumption that harmonics are Doppler shifted by varying
wave-induced wind directly to very short scales where they are removed by molecular diffusion. When
averaged over wave phases, this parameterization, however, yields the very same termination of
harmonics employing \textit{ad hoc} chosen criteria. The approach of \citet{MedvedevKlaassen95} is
based on the concept of ``enhanced diffusion" \citep{Weinstock76, Weinstock_etal07}. It takes into
account Doppler shift by larger-scale harmonics in the spectrum, and erosion by shorter-scale ones.
For parameterization purposes, Doppler shift can be neglected, the coefficient of eddy-induced
diffusion is self-consistently calculated, and no ``tuning parameters" are required
\citep{MedvedevKlaassen00}.

GW parameterizations suitable for thermosphere GCMs must account also for damping by
molecular diffusion, thermal conduction, and ion friction. This is usually done by
incorporating the respective dissipation terms into the complex dispersion relation in the
form of imaginary parts of frequencies.  The first parameterization of this kind has been
proposed by \citet{Matsuno82}, and the most recent derivation for molecular diffusion and
thermal conduction has been performed by \citet{VadasFritts05}.  This approach is based on
the assumption that dissipation is relatively weak, where the degree of ``weakness" depends
on the characteristics of the harmonic and the background flow. This constitutes
another limitation on GW parameterizations. Molecular viscosity grows exponentially with
height in the thermosphere, and, eventually, the dissipation terms can significantly exceed
all other balancing terms in the equations for waves.  This means that GWs degenerate into
other types (``viscous waves"), and can no longer be considered within the parameterization
framework.

We illustrate the principles outlined above, and discuss some general details of
implementation into a GCM using the extended GW 
parameterization \citep{Yigit_etal08}.

\subsection{The extended nonlinear spectral gravity wave parameterization}

The word ``extended" denotes that the parameterization has been extended to
account for wave propagation in the thermosphere in accordance with the
requirements outlined above \citep{YMbook13}. It solves the equation
for the vertical structure of the horizontal momentum flux (per unit mass)
$\overline{u^\prime w^\prime}$ associated with the harmonic $j$ from a given
spectrum of waves:
  \begin{equation}
    \frac{d\overline{u^\prime w^\prime}_j}{dz}= -\biggl( \frac{\rho_z}{\rho} 
    + \beta^j_{tot} \biggl) \overline{u^\prime w^\prime}_j,
    \label{eq:gwflx}
  \end{equation}
where $\beta^j_{tot}$ is the total vertical damping rate acting on the
harmonic. If propagation is conservative ($\beta^j_{tot}=0$), then the flux
$\rho \overline{u^\prime w^\prime}_j$ is constant with height. The total damping
rate for a given harmonic is the sum of the rates due to various dissipation
processes affecting the propagation and acting simultaneously
  \begin{equation}
    \beta^j_{tot} = \beta^j_{non} + \beta^j_{mol} + \beta^j_{ion} +
    \beta^j_{rad} + \beta^j_{eddy} + ...
    \label{eq:beta}
  \end{equation}

The main processes accounted for by the scheme include, correspondingly, nonlinear
breaking/saturation ($\beta^j_{non}$), molecular diffusion and thermal conduction ($\beta^j_{mol}$),
ion friction ($\beta^j_{ion}$), radiative damping ($\beta^j_{rad}$), and eddy diffusion
($\beta^j_{eddy}$) as suggested in the work by \citet{Yigit_etal08}. The term $\beta^j_{non}$ is
parameterized after the work by \citet{MedvedevKlaassen00}, and comprises the effects of other
harmonics on a given harmonic. Thus, the total wave field is not a simple collection of independent
waves, but of interacting ones.  The word ``nonlinear" in the name of the parameterization signifies
this.  Dissipation of a harmonic is strongly affected by changes in the background wind as the
  vertical damping is inversely proportional to the intrinsic phase speed of the harmonic, i.e.,
  $\beta^j\propto (c_j-u)^{-n}$, where the exponent $n$ differs for various dissipation mechanisms
  \citep[see e.g.,][]{Yigit_etal08,Yigit_etal09,Yigit_etal12b, YMbook13}. If the flux
$\rho \overline{u^\prime w^\prime}_j$ changes with height, the wave momentum is transferred to the
mean flow by means of an acceleration or deceleration, which is often called ``wave drag"
  \begin{equation}
    a_j=-\frac{1}{\rho}\frac{\mathrm{d}\rho \overline{u^\prime w^\prime}_j}{\mathrm{d}z}.
    \label{eq:acc}
  \end{equation}

The total ``drag" is determined by the gradient of the sum of fluxes for all
$M$ harmonics in the spectrum, $\Sigma_j^M a_j$. 

Equation (\ref{eq:gwflx}) is solved for each grid column of a GCM. For that, values of
$\overline{u^\prime w^\prime}_j$ must be specified at a certain height $z_s$ in the lower
atmosphere, which is considered as a source level.  This initialization is done in all GW
parameterizations, but the choice is extremely important for this scheme, because it
contains no other tuning parameters, and the source spectrum is the only input. A
representative spectrum can be seen in \citet[Figure 1,][]{Yigit_etal09}, where the fluxes
are specified as functions of horizontal phase velocities, and based on the observations of
\citet{Hertzog_etal08}.  The ``asymmetric" spectrum takes into account an anisotropy with
respect to the mean wind at the source level. The latter has been first suggested
heuristically \citep{Medvedev_etal98}, and a possible explanation has been offered recently
\citep{Kalisch_etal14}.  

GW harmonics with larger vertical wavelengths are less affected by dissipation and, therefore,
  tend to propagate higher. Typical scale height $H$ also increases in the thermosphere (e.g., $H$
  is around 50~km at 250 km altitude). Because the parameterization is based on the WKB
  approximation (section~\ref{sec:assumptions}), the vertical wavenumbers of accounted harmonics are
  limited by the relation $k_zH \gg 1$.  This relation translates into the limitation on the
    maximum phase velocities of GW harmonics considered in the parameterization to be 80 to
  100 m~s$^{-1}$.

Using general circulation modeling, the extended GW scheme has been extensively
  validated against the empirical horizontal wind model (HWM) \citep{Yigit_etal09} and the
  MSIS temperature distributions \citep{YigitMedvedev09}.  In a planetary atmospheres
context, the extended scheme has successfully been used in a state-of-the art Martian GCM in
order to investigate GW-induced dynamical and thermal coupling processes
\citep{Medvedev_etal13, MedvedevYigit12, Yigit_etal15a, Medvedev_etal16}.

\section{Effects of internal gravity waves on the general circulation of 
the upper atmosphere} 
\label{sec:gweffects}

Given the statistical approach to parameterizing waves, in which all the information on wave
phases is lost, and given the set of assumptions listed in
section~\ref{sec:assumptions}, no effects of individual wave packets can be
simulated with GCMs. They can only be approached with GW-resolving models similar to that of
\citet{Miyoshi_etal14}.  Historically, the need for accounting for 
GW effects emerged from an inability of GCMs to reproduce the observed zonal mean
circulation in the middle atmosphere \citep{Holton83}.  In particular, the inclusion of
parameterized effects of subgrid-scale waves has helped to realistically simulate the
semi-annual oscillation in the MLT (mesosphere and lower thermosphere) with a GCM
\citep{MedvedevKlaassen01}.  \citet{Manson_etal02} demonstrated the same for solar tides.
Recently, \citet{Schirber_etal14} have shown that, with the use of a convection-based GW
scheme, a GCM has reproduced a quasi-biennial oscillation (QBO) with realistic features.

 \begin{figure}[t!]
\includegraphics[width=0.45\textwidth]{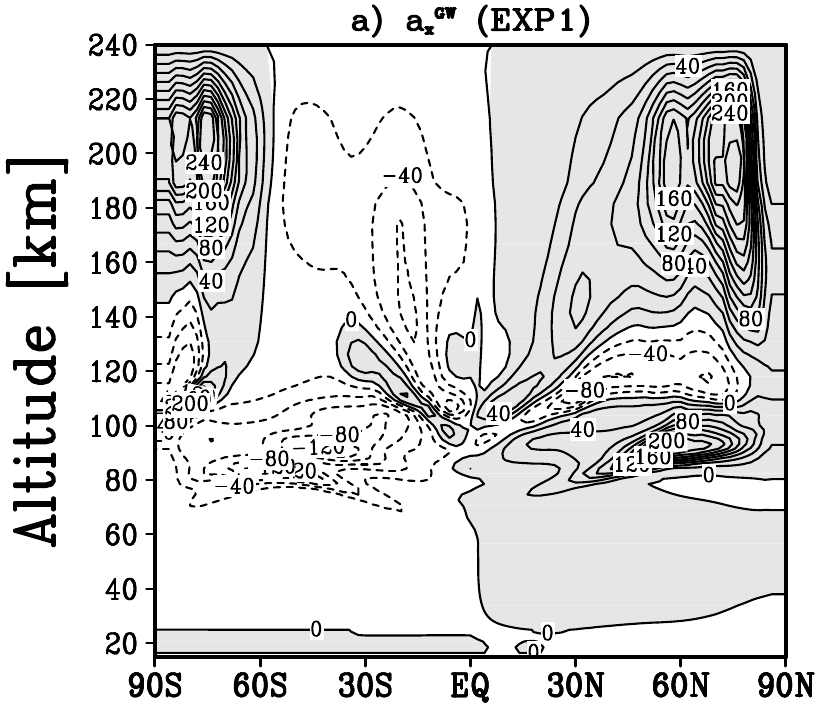}
\caption{Parameterized gravity wave drag. Altitude-latitude 
distribution of the parameterized zonal-mean zonal gravity wave drag 
(in m~s$^{-1}$~day$^{-1}$ averaged over June/July solstice conditions 
based on the simulation with the CMAT2 GCM incorporating the whole atmosphere paramaterization of \citet{Yigit_etal08}. After \citet[Figure 10,][]{YigitMedvedev10}.}
\label{fig:Y09_f3}
 \end{figure}

Studying the effects of GWs of tropospheric origin in the thermosphere has a long history
\citep[see][for more detail]{YigitMedvedev15}, however their dynamical importance at higher
altitudes has not been fully recognized until recently.  In all GCMs extending into the
thermosphere, the effects of subgrid-scale GWs were either neglected, or assumed to decay
exponentially above a certain height (e.~g., turbopause $\sim105$ km).  Simulations of
\citet{Yigit_etal09} with the Coupled Middle Atmosphere and Thermosphere-2
\citep[(CMAT2),][]{Yigit09} GCM incorporating the extended nonlinear parameterization
of \citet{Yigit_etal08} revealed that the momentum deposition by lower atmospheric GWs in
the F region is substantial, and is comparable to that by ion drag.  Figure~\ref{fig:Y09_f3}
shows the latitude-altitude distribution of the simulated zonal mean zonal forcing by
parameterized GWs. This forcing (known as ``GW drag") is directed mainly
against the mean zonal wind, and plays an important role in the momentum balance of the
upper thermosphere, similar to the scenario in the middle atmosphere. The magnitude of
thermospheric GW drag, exceeding $\pm 200$ m s$^{-1}$ day$^{-1}$, is larger than its effects
in the middle atmosphere.

\begin{figure}[t!]
 \includegraphics[width=0.45\textwidth]{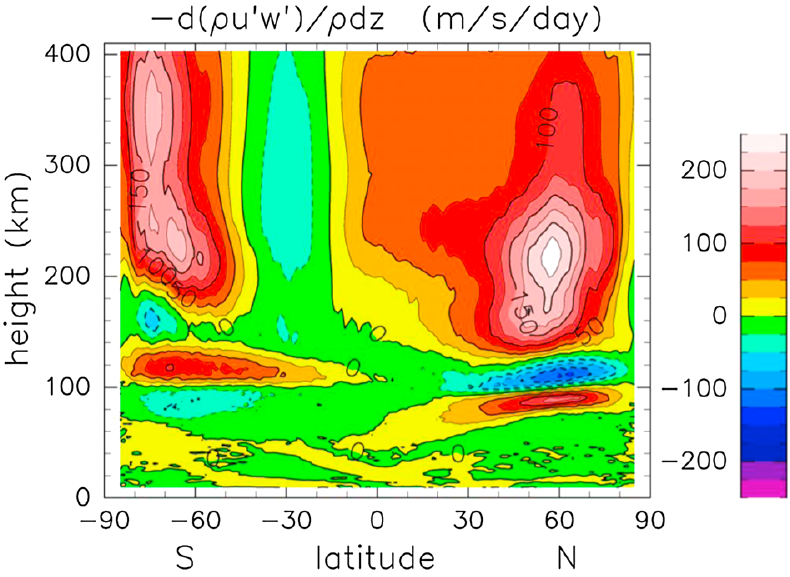}
  \caption{Modeled zonal gravity wave drag. Same as in
    Figure~\ref{fig:Y09_f3}, but for the drag due to explicitly
    resolved gravity waves in simulations with the GAIA GCM.
    After \citet[Figure 3,][]{Miyoshi_etal14}.}
  \label{fig:miyoshi}
\end{figure}

\citet{Miyoshi_etal14}'s recent simulations with a whole atmosphere GW-resolving GCM have confirmed
\citet{Yigit_etal09}'s predictions of the appreciable dynamical effects of lower atmospheric GWs on
the general circulation of the thermosphere above the turbopause. Figure~\ref{fig:miyoshi} presents
the divergence of momentum fluxes ($a$ in equation \ref{eq:acc}) due to the resolved portion of GW
spectra (with horizontal scales longer than 380 km) calculated for solstice conditions
\citep[][Figure 3]{Miyoshi_etal14} as in the GCM modeling by \citet{Yigit_etal09}. Considering the
various approximations and limitations of the extended parameterization, and, especially,
uncertainties with specifying GW sources, the two distributions in Figures~\ref{fig:Y09_f3} and
\ref{fig:miyoshi} appear to be in a good qualitative and quantitative agreement.  There are also
some differences between the two simulations. In particular, in the Southern Hemisphere MLT, the
high-resolution simulations show a region of eastward GW, which is only present at the Southern
Hemisphere high- and low-latitudes in the parameterized simulation. Two possible sources of the
discrepancies are the source spectrum and effects of the background winds on the propagation and the
resulting dissipation. Overall, both simulation studies demonstrated that, due to propagation
conditions in the middle atmosphere, most of the thermospheric GW activity concentrate at
high-latitudes, where solar tides modulate local time variations of GW drag. This, and further
analyses of the simulations with the high-resolution model provided evidences that thermospheric
effects of GWs can be successfully parameterized in lower-resolution GCMs.

Thermal effects of GWs are two-fold: (a) heating due to conversion of the 
mechanical energy of dissipating harmonics into heat, and (b) heating and 
cooling associated with the downward sensible heat flux 
$\overline{w^\prime T^\prime}$ induced by these waves
\citep{MedvedevKlaassen03,Becker04}. Magnitudes of the former in the 
thermosphere are comparable with those due to the Joule heating, while the 
latter is comparable with the cooling rates due to molecular thermal conduction 
\citep{YigitMedvedev09}, which suggests that the thermal effects of 
GWs cannot be neglected in the upper atmosphere.
\citet{YigitMedvedev10}'s GCM simulations with the extended scheme have 
demonstrated that the variations of thermospheric GW effects are appreciable.  
GWs propagate to higher altitudes during high solar activity, but produce 
weaker drag than during periods of low solar activity. Their observations have 
later been qualitatively verified by the satellite observations of 
\citet{Park_etal14}.



\section{Sudden Stratospheric Warmings}
\label{sec:ssw}
\subsection{Characteristics}
\label{ssw_character}

Sudden stratospheric warmings first discovered observationally by \citet{Scherhag52} are
transient events during which the eastward zonal mean zonal winds weaken, or even reverse
the direction at 60$^\circ$N (geographic) at $\sim 30$ km (10 hPa), followed by the
significant warming of the winter North Pole (90$^\circ$N) \citep{Labitzke81,
  andrews_etal87}. Since the 1950s, as the interest in studying SSWs has grown, the
  classification of SSWs has evolved \citep[see][for a comprehensive
  discussion]{Butler_etal15}. Essentially, there are two commonly accepted types of
  warmings: a minor and a major warming.  The warming is \textit{major} if the
equator-to-pole temperature gradient reverses poleward of 60$^\circ$ latitude in addition to
the reversal of the zonal mean zonal winds at 60$^\circ$N at 10 hPa
  \citep{Labitzke81}. If the westerly mean zonal wind weakens but does not reverse the
direction, i.e., the stratospheric vortex does not break
  down, during a temperature increase at the Pole, then the warming is defined as a
\textit{minor} event.

\begin{figure}[t!]
  \includegraphics[width=0.45\textwidth]{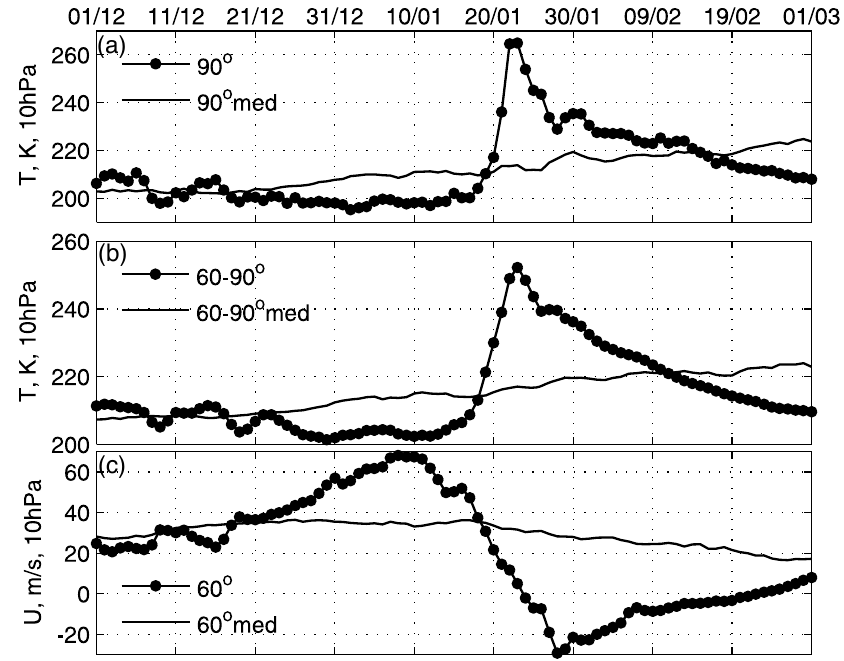}
  \caption{The 2008--2009 sudden stratospheric warming.
    Variation of the
    stratospheric conditions at 10 hPa during the sudden stratospheric warming
    that took place in the winter of 2008--2009 according to data from the
    National Center for Environmental Predictions (NCEP).
    From top to bottom: Stratospheric
    temperature at 90$^\circ$N); mean temperature at 60$^\circ$-90$^\circ$N;
    mean zonal wind at 60$^\circ$N. Thin lines represent 30-year means of
    stratospheric parameters. 
    Adopted from \citet[Figure 1]{Goncharenko_etal10}. }
  \label{fig:ssw}
\end{figure}
An illustration of the major SSW features is seen in Figure~\ref{fig:ssw} for 
a representative major warming that took place in the winter of 2008--2009, as
adopted from the work by \citet[Figure 1]{Goncharenko_etal10}. These
stratospheric conditions are based on data from the National Center for
Environmental Prediction (NCEP). Within about 5 days, the zonal mean 
temperature at 10 hPa increases by more than 60 K (from 200 to more than $260$ 
K) at the North Pole, that is, more than $30\%$ increase (top panel). 
The average temperature at high-latitudes (60$^\circ$ to 90$^\circ$N) 
increases significantly as well. The eastward (positive) zonal mean zonal wind 
starts decelerating already before the onset of the warming at the Pole and, 
reverses its direction, reaching a minimum over a period of about 10 days 
(bottom panel). The thin solid curves in each panel show the 30-year means of 
the associated parameters. \citet{Goncharenko_etal10} have also demonstrated 
in their analysis that the 2008-2009 warming was related to a weakening of the 
planetary wave-1 and an enhancement of the wave-2.

A comprehensive review of the earlier theoretical explanations of SSWs can be found in the
works by \citet{Schoeberl78} and \citet{Holton80}. Earlier studies have indicated that
planetary-scale waves have to be properly taken into account during warming
periods. According to the seminal work of \citet{CharneyDrazin61}, planetary-scale
disturbances can propagate from the troposphere into the stratosphere in the presence of
prevailing westerlies, and the transport of eddy heat and momentum by vertically propagating
waves is expected to modify the stratospheric zonal flow. Initial idealized simulations of
wave propagation have suggested that planetary waves with wave numbers 1 and 2 can reach the
stratosphere \citep{Matsuno70}.  \citet{Matsuno71} modeled that Rossby wave-mean flow
interactions decelerate the polar night jet, leading to weakening and even breakdown of the
polar vortex, and ultimately to a sudden warming of the polar region. Later, the numerical
works by \citet{Holton76} and \citet{Palmer81} have qualitatively provided supporting
evidence for \citet{Matsuno71}'s model.

\subsection{Mechanism of the sudden warming}

In the winter (solstice) period, the Northern Hemisphere stratosphere is dominated by
westerly jets whose strength increases with altitude. Quasi-stationary planetary waves can
propagate vertically upward, provided that the mean zonal flow satisfies the conditions for
vertically propagating wave modes. For these waves, the zonal wind has to fulfil the
following condition \citep[][Equation (12.16)]{HoltonHakim12}:
  \begin{equation}
    \label{eq:uc1}
    0 < \bar{u} < u_c,
  \end{equation}
where the Rossby critical velocity $u_c$ is defined in terms of the
characteristics of the background atmosphere and wave by 
  \begin{equation}
    \label{eq:uc2}
    u_c \equiv \beta\, k_h^2 + \frac{f_0^2}{4\, N^2\, H^2},   
  \end{equation}

where $k_h^2 = k^2 + l^2$ is the horizontal wavenumber that depends on the zonal
($k=2\pi/\lambda_x$) and the meridional ($l=2\pi/\lambda_y$) wavenumbers; $f = f_0 - \beta y$ is the
beta-plane approximation for the Coriolis parameter, and
$\beta \equiv \frac{\partial f}{\partial y}$ is the meridional gradient of the Coriolis
parameter. The condition (\ref{eq:uc1}) suggests that planetary waves can propagate vertically only
in the presence of westerly winds that are weaker than a certain critical value $u_c$, which depends
on the horizontal scale of the wave. Dynamical conditions are, therefore, favorable for the vertical
propagation of planetary waves in the winter Northern Hemisphere with prevalent mean westerly
winds. This condition is important for understanding the propagation of GWs, which are also affected
by the mean wind distributions. Namely, before the warming, the stratospheric zonal mean winds are
eastward. They filter out a significant portion of the eastward directed GWs, favoring the upward
propagation of harmonics with phase velocities directed westward.  During the warmings, the
decelerating westerlies increase the chances of GWs with eastward horizontal phase speeds to
propagate to higher altitudes \citep{YigitMedvedev12}.

In the winter stratosphere, waves are rapidly attenuated, thus decelerating the mean zonal
flow. For the occurrence of SSWs, a large-scale wave transience, in particular, rapid temporal
  changes of planetary wave activity are also important.  They maintain the convergence of the
  westward momentum flux, leading to strong polar night jet deceleration and poleward meridional
  flow enhancement \citep{andrews_etal87}.  Additionally, radiative forcing sustains a cold winter
North Pole with negative equator-to-pole mean temperature gradient, that is,
$\frac{\partial T}{\partial y} < 0$. The rapid deceleration of the stratospheric mean flow implies a
decreasing (positive) vertical gradient of the zonal flow between the troposphere and
stratosphere. From the thermal wind relation
$\frac{\partial u}{\partial z} \sim - \frac{\partial T}{\partial y}$, this decrease implies a rise
of temperature at the winter pole, meaning that the equator-to-pole mean temperature gradient
becomes less negative. During a major warming, this gradient even reverses due to the reversal of
the vertical gradient of zonal mean wind. The strong polar night jet deceleration leads to a
departure from the thermal wind balance, and the poleward meridional flow, which is caused by
  the Coriolis force associated with the westward forcing, is induced to recover this balance.
This enhancement of the Brewer-Dobson circulation ultimately results in an adiabatic warming at
Northern Hemisphere high-latitudes.

\section{Observed changes in the upper atmosphere 
during sudden stratospheric warmings}
\label{sec:ssw-observed}

Given the rapid and strong local changes in the circulation and thermal structure of the
stratosphere during SSWs, the natural questions that bear in mind are (1) how high the
effects of the warming propagate in altitude, and (2) to what extent the changes in the
upper atmosphere can be associated with the sudden warmings. Planetary waves 
 cannot
propagate directly to much higher altitudes, but the stratosphere and mesosphere are closely
connected via circulation and by GWs and tides. As sudden warmings and the associated
dynamical changes in the stratosphere occur over relatively long time scales (e.g.,
$\sim 10$ days) compared to the periods of internal waves, lower atmospheric wave
disturbances have sufficient time to propagate to higher altitudes, provided that
propagation conditions are favorable. Therefore, one ought to expect a certain degree of
coupling between the stratosphere and higher altitudes, probably beyond the middle
atmosphere.

\begin{figure*}
  \begin{center}
    \includegraphics[width=0.85\textwidth]{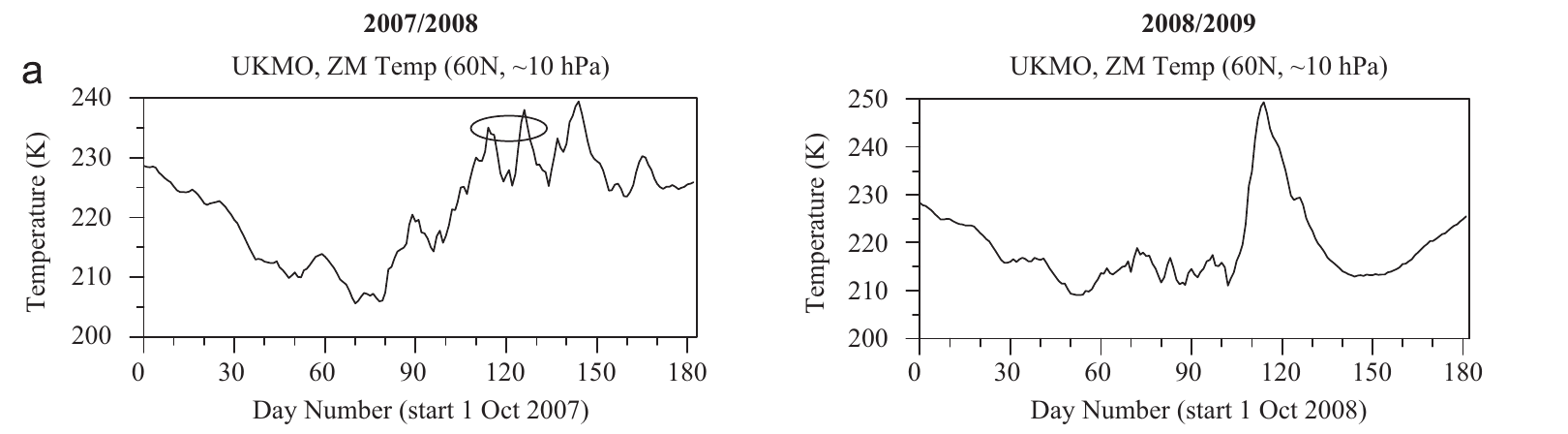}

    \includegraphics[width=0.85\textwidth]{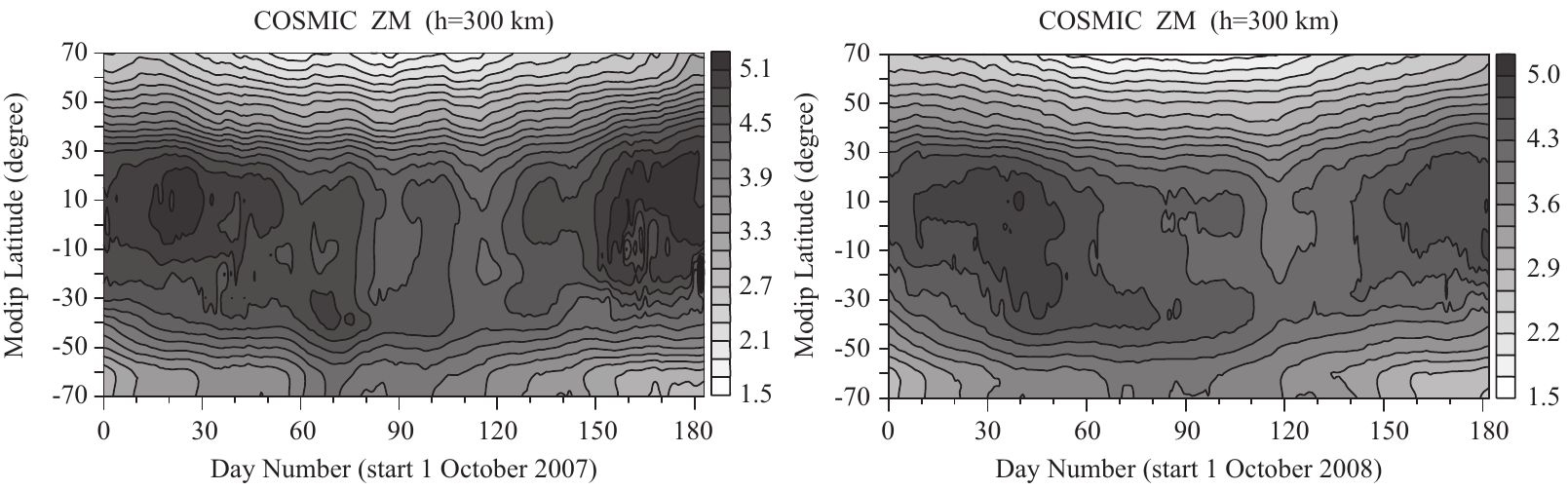}
    \caption{SSW event and ionospheric variations. The upper panels show the 2007/2008 (left) and
      2008/2009 (right) SSWs according to the UK Met Office (UKMO) model. In the lower panels,
      the associated variations of the mean zonal mean electron density (in MHz) retrieved from
      COSMIC at 300 km during the two warmings are seen.  Adopted from \citet[Figures 1,
      2]{PanchevaMukhtarov11}.}
  \end{center}

  \label{fig:Pancheva}
\end{figure*}

How can one associate observed upper atmospheric changes with SSWs? Essentially, a
ground-to-upper atmosphere observation with a single instrument is beyond the capabilities
of the current technology. For the purposes of observational analysis, SSW events/periods
ought to be identified.  For this, an appropriate description of stratospheric dynamics is
needed in the first place. This representation could be, for example, obtained from
numerical forecast models that  assimilate in-situ and remote-sensing data, such
as the European Centre for Medium-Range Weather Forecast (ECMWF) analyses, and produce the
required global fields of atmospheric parameters. Then, observational data can be
investigated together with the numerical model output \citep[e.g.,][]{Pancheva_etal08}.

The deceleration of the stratospheric eastward zonal flow during sudden warmings
leads, ultimately, to an upward circulation in the mesosphere that results in
mesospheric cooling \citep{LiuRoble02}. Such direct link between these two
regions have motivated a number of scientists to investigate the details of
stratosphere-mesosphere changes during warmings. Based on temperature and
geopotential height data obtained from the Sounding the Atmosphere using
Broadband Emission Radiometry (SABER) instrument of the Thermosphere Ionosphere
Mesosphere Energetics and Dynamics (TIMED) satellite and the VHF radar
horizontal winds, \citet{Pancheva_etal08} have investigated planetary
wave-induced coupling in the stratosphere-mesosphere during the major warming 
of 2003/2004 winter Northern Hemisphere.

\citet{Yuan_etal12} studied the response of the middle-latitude mesopause region to the 2009 major
SSW, using a sodium Doppler wind-temperature lidar.  They have discovered anomalous behavior of the
mean temperature and zonal winds around the mesopause during the warming, and have concluded that it
was due to a direct impact of the major warming on the middle-latitude mesopause. The 2009 SSW has
been one of the strongest warming events that has been recorded. The features around the mesopause
during SSWs can be largely characterized in terms of an ``elevated stratopause", which forms around
75--80 km after the SSW occurrence and then descends \citep{Maney_etal09}. The role of GWs and
planetary-scale waves in the time evolution of the elevated stratopause have been investigated by a
number of authors \citep[e.g.,][]{Siskind_etal10, Chandran_etal11,Limpasuvan_etal12}.

Vertical coupling between the stratosphere and the lower thermosphere has been studied in
the low- and middle-latitude Northern Hemisphere winter of 2003/2004 based on the
temperature data from SABER/TIMED \citep{Pancheva_etal09}.
According to \citet{GoncharenkoZhang08}'s analysis of the Millstone Hill incoherent scatter
radar (ISR) ion temperatures data, warming in the lower thermosphere and cooling above 150
km were revealed during a minor SSW. Using data from the Jicamarca ISR, \citet{Chau_etal09}
have detected significant semidiurnal tidal variations in the vertical
${\bf E}\times {\bf B}$ ion drifts in the equatorial ionosphere during the winter 2007-2008
minor warming. Using temperature measurements from the Michelson Interferometer for Passive
Atmospheric Sounding (MIPAS) on board European Space Agency's (ESA) Envisat satellite
measurements, \citet{Funke_etal10} have demonstrated observational evidence for the
dynamical coupling between the lower and upper atmosphere during the 2009 major SSW. Based
on TEC (total electron content) data retrieved from a worldwide network of GPS observations,
\citet{Goncharenko_etal10} have found a significant local time modulation of the equatorial
ionization anomaly (EIA) induced by SSWs. Using the European Incoherent Scatter (EISCAT) UHF
radar, \citet{Kurihara_etal10} have detected short-term variations in the upper atmosphere
during the 2009 major warming. In their analysis of Fabry-Perot and incoherent scatter radar
data, \cite{CondeNicolls10} have identified that the period of reduced neutral temperatures
at 240 km, which corresponded closely to the main phase of the warming.

More recently, analyzing the Constellation Observing System for Meteorology Ionosphere and
Climate (COSMIC) data, \citet{PanchevaMukhtarov11} have found a systematic negative response
of ionospheric plasma parameters ($f_0F_2$, $h_mF_2$, and $n_e$) during the warmings of
2007--2008 and 2008--2009.  An illustration of their results for the mean zonal mean
electron density (in MHz) at 300 km are seen in Figure~\ref{fig:Pancheva}, where the
2007--2008 and 2008--2009 warming events are shown on the left and right panels,
respectively. The response to the warming is negative and mainly occurs in the low- and
middle-latitude region. \citet{Liu_etal11} used neutral mass density observations from the
CHAMP (Challenging Minisatellite Payload) and GRACE (Gravity Recovery and Climate
Experiment) satellites to study the thermospheric variations during the 2009 major
warming. They have found a substantial decrease of the mass density, and concluded that it
was potentially associated with a strong thermospheric cooling of about 50 K.
\citet{Goncharenko_etal13} have investigated the day-to-day variability in the
middle-latitude ionosphere during the 2010 major warming using the Millstone Hill ISR. They
have discovered that semidiurnal and terdiurnal tidal variations were enhanced during the
SSW. \citet{Jonah_etal14} have used a suit of observational data from GPS, magnetosphere,
and meteor radar in order to investigate the response of the magnetosphere and ionosphere to
the 2013 major SSW. Analyzing long-term data of the global average thermospheric total mass
density derived from satellite orbital drag, \citet{Yamazaki_etal15b} showed density
reduction of 3–7 \% at 250–575 km during SSW period that can be associated with lower
atmospheric forcing. Recently, using data from GPS and ionosonde stations,
\citet{Fagundes_etal15} investigated the response of the low- and middle-latitude ionosphere
in the Southern Hemisphere to the 2009 major SSW and found that during the warming, TEC was
depressed following the SSW temperature peak.


Overall, these studies suggest that (1) a variety of instruments has been used to study
upper atmospheric changes during SSWs; (2) They convincingly demonstrated that SSW events
affect the thermosphere-ionosphere system beyond the turbopause; and (3) the associated
observed changes in the upper atmosphere vary from one warming event to another. Some
studies indicate that large-scale internal wave processes may be involved in vertical
coupling during SSWs. One of the less appreciated topics in SSW studies is the role of
small-scale GWs.  We next discuss the upper atmosphere changes during SSWs in the context of
lower atmospheric small-scale GWs that can propagate to the thermosphere
\citep{Yigit_etal09, Yigit_etal12b}.

\section{Upper atmospheric changes during sudden stratospheric warmings }
\label{sec:gwssw}

\begin{figure}[h!]
  \includegraphics[width=0.45\textwidth]{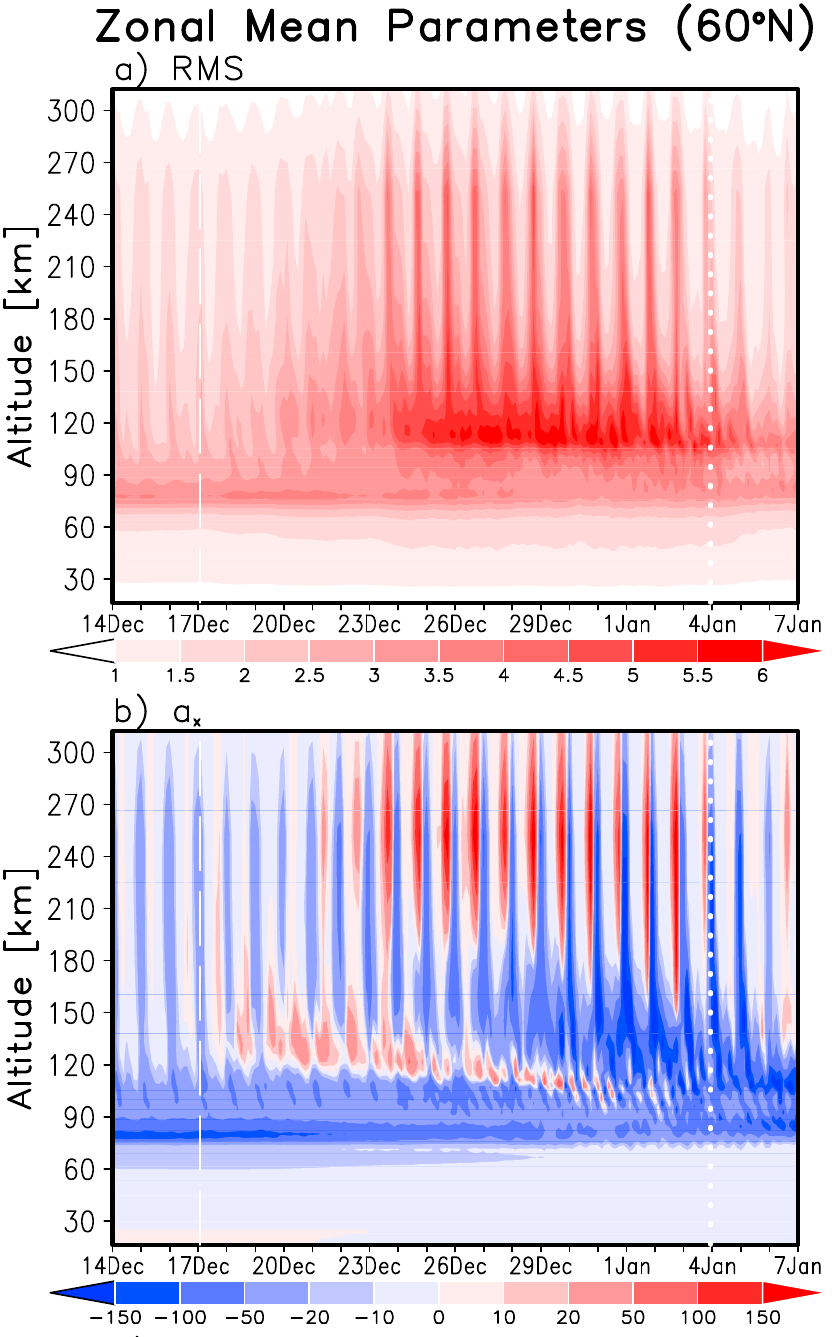}
  \caption{GW-induced root-mean-square (RMS) wind and GW drag UT
      variations during a minor SSW. Altitude-UT variations of the zonal mean
    (a) GW-induced RMS wind fluctuations in m s$^{-1}$ and (b) zonal drag in m
    s$^{-1}$ day$^{-1}$, where red (positive) is eastward and blue (negative) is westward
    drag. White vertical dashed and dotted lines denote the onset
    and the end of the minor sudden warming \citep[Figure 2]{YigitMedvedev12}.}
  \label{fig:gw_ssw_rms_wind}
\end{figure}

Observing dynamical changes, e.g., with satellites and radars,  cannot provide sufficiently
detailed information on characteristics and physical mechanisms of vertical
coupling. Observations may, and in fact do raise new questions, which can be investigated by
models.  A powerful tool is general circulation models (GCMs) that solve the coupled
governing equations of atmospheric and ionospheric physics in time and three-dimensional
space.  GCMs generate a full set of field parameters that can be diagnosed in detail in
order to investigate the physical mechanisms that shape the state and evolution of the
atmosphere. Therefore, global models can provide an unprecedented insight in vertical
coupling processes between the different atmospheric regions. One should nevertheless be
aware of the limitations of GCMs, such as,  resolution, boundary conditions, and
parameterizations.


As discussed in section \ref{sec:gweffects}, GWs have a profound effect on the dynamical
\citep{Yigit_etal09, Yigit_etal12b, Miyoshi_etal14, Vadas_etal14}, thermal
\citep{YigitMedvedev09,Hickey_etal11} and compositional \citep{WalterscheidHickey12} structure of
the upper atmosphere. The state of the background middle atmosphere plays a crucial role in
modulating the propagation of GWs into the thermosphere.  Given that SSWs modify the middle
atmospheric circulation significantly, how can they influence the upper atmosphere in the context of
GW propagation and dissipation in the thermosphere? Resolving this science question requires a use
of comprehensive GCMs with appropriate representation of small-scale GWs. The general circulation
modeling of \citet{YigitMedvedev12} using the extended GW parameterization of \citet{Yigit_etal08}
has demonstrated GW propagation and appreciable dynamical effects in the upper thermosphere during a
minor warming. The universal time (UT) variations of the GW-induced zonal mean root-mean-square
(RMS) wind fluctuations (in m~s$^{-1}$) and zonal mean GW drag (in m s$^{-1}$~day$^{-1}$) are shown
in Figures~\ref{fig:gw_ssw_rms_wind}a,b.  GW-induced RMS wind fluctuations are given by
  $\sigma = \big( M^{-1}\sum_j^M\overline{u_j^{\prime 2}}\big)^{1/2} $, where $M $ is the number of
  harmonics in the spectrum and variance $\overline{u_j^{\prime2}}$ is related to the wave amplitude
  as $\sqrt{\overline{u_j^{\prime2}}} \equiv |u_j^\prime|$.  The GW RMS wind fluctuations are
a proxi for the subgrid-scale GW activity as the fluctuations induced by all waves in a GW
spectrum are taken into account and do not represent the resolved wind fluctuations.  In the
course of the warming, GW activity increases by a factor of three exceeding 6 m~s$^{-1}$ in response
to weakening of the polar night jet.  In addition to persistently large values in the lower
thermosphere, modulation of the GW activity is seen higher in the thermosphere. Following the
increase of GW activity, (eastward) GW drag increases in the thermosphere during the warming as
well.

The effects of GWs in the upper atmosphere during SSWs are not confined to only those in a zonal
mean sense. Recently, \citet{Yigit_etal14} have investigated the details of GW temporal variations
in the thermosphere during a minor warming simulated with a GCM. They modeled that GW drag and its
variability is dramatically enhanced in the thermosphere during the warming and thus lead to a
$\sim\pm50\%$ modulation of small-scale and short-term variability in the resolved thermospheric
winds, where the small-scale variability has been evaluated by subtracting the contributions of the
large-scale tides. Recently, \cite{Miyoshi_etal15} have demonstrated with a GW-resolving GCM that
the SSW has major dynamical and thermal impact on the upper atmosphere, substantially influencing
the global circulation.  Changes in the mean zonal wind produce a feedback on GWs by modifying
filtering, dissipation, and propagation conditions.

The upper atmosphere above the turbopause has a great amount of variability 
owing to the simultaneous influences of meteorological and space weather 
processes \citep{Matsuo_etal03,Anderson_etal11,YigitRidley11b,Yigit_etal12a}. 
Often, separating the components and sources of variability in observations is 
a challenging task. Thus, following their observations of an SSW, 
\citet{Kurihara_etal10} have concluded that understanding the link between 
SSWs and thermal and dynamical changes in the upper atmosphere-ionosphere 
requires investigations of GW-mean flow interactions processes. GCM studies 
can greatly supplement these efforts.

Predictions with GCMs indicate that small-scale GWs can substantially contribute
to the variability of the upper atmosphere. Also, recent modeling studies with a
whole-atmosphere GCM have shown an enhancement of the semidiurnal variation in
the ionospheric ${\bf E}\times {\bf B}$ drifts during the 2009 major warming
\citep{Jin_etal12}. This increase has been interpreted as a consequence of the
semidiurnal tidal amplification in the lower atmosphere. 
Further investigations that
incorporate a fully two-way coupled thermosphere-ionosphere under the influence
of lower atmospheric waves are required in order to assess the significance of
the lower atmosphere in the context of upper atmosphere variability. In
characterizing the upper atmosphere processes, the variability is always defined
with respect to some appropriate mean.  Therefore, the quantity of variability
is not uniquely defined, and care should be taken when comparing one study to
another. In a broader context, the presence of any kind of variability restricts
scientists' ability to predict the future state of the atmosphere, and it is
crucial to determine the sources of variability and quantify the magnitude
thereof.

\section{Conclusions}
\label{sec:conc}
This paper has briefly reviewed the
current state of knowledge and most recent developments with understanding the
role of GWs in vertical coupling during SSWs. The observed upper atmosphere
changes during SSWs have been presented. An emphasis was placed on the 
processes above the mesopause, and on how they can be studied with general
circulation models.

The geosciences community increasingly recognizes that the effects of lower
atmospheric gravity waves extend beyond the middle atmosphere into the
atmosphere-ionosphere system and are of global nature.  Similarly, sudden
stratospheric warmings were used to be looked upon as stratospheric phenomena,
but now compelling observational evidences of their signatures in the
thermosphere-ionosphere are being routinely provided.

With the rapid progress in the field of atmospheric coupling, further key
science questions on the role of GWs in coupling atmospheric layers arise:
\begin{itemize}
\item  What are the spectra of gravity waves in the lower and upper atmosphere?
  How do they change during the different phases of SSWs?
\item How well do GW parameterizations describe wave spectra and reproduce their
  effects during SSWs?
\item What is the relative role of GW- and electrodynamical coupling 
  between atmospheric layers in the variability of the atmosphere-ionosphere 
  system during SSWs?
\item What are the effects of GWs on the circulation 
  and thermal
  budget of the upper atmosphere during major sudden stratospheric warmings?
\item Do GWs in the upper atmosphere affect the development of sudden
  stratospheric warmings, or they are a mere reflection of processes in the
  lower atmosphere?
\item Do GWs have a role in latitudinal coupling in the thermosphere
 during SSW events?
\end{itemize}
This is certainly an incomplete list of scientific questions, answering which
requires concerted observational, theoretical, and modeling efforts from
scientists of both lower and upper atmosphere communities.

 \section*{Competing interests}
   The authors declare that they have no competing interests.

 \section*{Author's contributions}
 Both authors have participated in writing all sections, read and approved the final manuscript. 

\acknowledgements{
 The work was partially supported by German Science Foundation (DFG) grant
 ME2752/3-1. EY was partially supported by NASA grant NNX13AO36G and NSF grant AGS 1452137.}

\bibliography{mybib-v36}

\begin{thebibliography}{}
\expandafter\ifx\csname natexlab\endcsname\relax\def\natexlab#1{#1}\fi

\bibitem[{Anderson {et~al.}(2011)Anderson, Davies, Conde, Dyson, \&
  Kosch}]{Anderson_etal11}
Anderson, C., Davies, T., Conde, M., Dyson, P., \& Kosch, M.~J. 2011, J.
  Geophys. Res., 116

\bibitem[{Andrews {et~al.}(1987)Andrews, Holton, \& Leovy}]{andrews_etal87}
Andrews, D.~G., Holton, J.~R., \& Leovy, C.~B. 1987, International geophysics
  series, Vol.~40, Middle Atmosphere Dynamics (Academic press)

\bibitem[{Becker(2004)}]{Becker04}
Becker, E. 2004, J. Atmos. Sol.-Terr. Phys., 66, 683

\bibitem[{Becker(2011)}]{Becker11}
---. 2011, Space Sci. Rev., 168, 283

\bibitem[{Beres {et~al.}(2004)Beres, Alexander, \& Holton}]{Beres_etal04}
Beres, J.~H., Alexander, M.~J., \& Holton, J.~R. 2004, J. Atmos. Sci., 61, 324

\bibitem[{Butler {et~al.}(2015)Butler, Seidel, Hardiman, Butchart, Birner, \&
  Match}]{Butler_etal15}
Butler, A.~H., Seidel, D.~J., Hardiman, S.~C., {et~al.} 2015, Bull. Amer.
  Meteorol. Soc., 96, 1913, 2014JE004715

\bibitem[{Chandran {et~al.}(2011)Chandran, Collins, Garcia, \&
  Marsh}]{Chandran_etal11}
Chandran, A., Collins, R.~L., Garcia, R.~R., \& Marsh, D.~R. 2011, Geophys.
  Res. Lett., 38, doi:10.1029/2010GL046566

\bibitem[{Charney \& Drazin(1961)}]{CharneyDrazin61}
Charney, J.~G., \& Drazin, P.~G. 1961, J. Geophys. Res., 66, 83

\bibitem[{Charron \& Manzini(2002)}]{CharronManzini02}
Charron, M., \& Manzini, E. 2002, J. Atmos. Sci., 59, 932

\bibitem[{Chau {et~al.}(2009)Chau, Fejer, \& Goncharenko}]{Chau_etal09}
Chau, J.~L., Fejer, B.~G., \& Goncharenko, L.~P. 2009, Geophys. Res. Lett., 36,
  doi:10.1029/2008GL036785

\bibitem[{Chun \& Baik(2002)}]{ChunBaik02}
Chun, H.-Y., \& Baik, J.-J. 2002, J. Geophys. Res., 59, 1006

\bibitem[{Conde \& Nicolls(2010)}]{CondeNicolls10}
Conde, M.~G., \& Nicolls, M.~J. 2010, J. Geophys. Res. Atmos., 115,
  doi:10.1029/2010JD014280, d00N05

\bibitem[{Fagundes {et~al.}(2015)Fagundes, Goncharenko, de~Abreu, Venkatesh,
  Pezzopane, de~Jesus, Gende, Coster, \& Pillat}]{Fagundes_etal15}
Fagundes, P.~R., Goncharenko, L.~P., de~Abreu, A.~J., {et~al.} 2015, J.
  Geophys. Res. Space Physics, 120, 7889, 2014JA020649

\bibitem[{Fritts \& Alexander(2003)}]{FrittsAlexander03}
Fritts, D.~C., \& Alexander, M.~J. 2003, Rev. Geophys., 41,
  doi:10.1029/2001RG000106

\bibitem[{Fritts \& Lund(2011)}]{FrittsLund11}
Fritts, D.~C., \& Lund, T.~C. 2011, in Aeronomy of the Earth's Atmosphere and
  Ionosphere, IAGA Special Sopron Book Series (Springer Netherlands), 109--130

\bibitem[{Funke {et~al.}(2010)Funke, L\'{o}pez-Puertas, Bermejo-Pantale\'{o}n,
  Garcia-Comas, Stiller, von Clarmann, Kiefer, \& Linden}]{Funke_etal10}
Funke, B., L\'{o}pez-Puertas, M., Bermejo-Pantale\'{o}n, D., {et~al.} 2010,
  Geophys. Res. Lett., 37, doi:10.1029/2010GL043619

\bibitem[{Garcia \& Solomon(1985)}]{GarciaSolomon85}
Garcia, R.~R., \& Solomon, S. 1985, J. Geophys. Res., 90, 3850, implementation
  of Lindzen's parameterization into a two-dimensional dynamical model to study
  the effects of GWs in the MLT

\bibitem[{Geller {et~al.}(2013)Geller, Alexander, Love, Bacmeister, Ern,
  Hertzog, Manzini, Preusse, Sato, Scaife, \& Zhou}]{Geller_etal13}
Geller, M., Alexander, M.~J., Love, P.~T., {et~al.} 2013, J. Clim., 26, 6383

\bibitem[{Goncharenko \& Zhang(2008)}]{GoncharenkoZhang08}
Goncharenko, L., \& Zhang, S.-R. 2008, Geophys. Res. Lett., 35,
  doi:10.1029/2008GL035684

\bibitem[{Goncharenko {et~al.}(2010)Goncharenko, Coster, Chau, \&
  Valladares}]{Goncharenko_etal10}
Goncharenko, L.~P., Coster, A.~J., Chau, J.~L., \& Valladares, C.~E. 2010, J.
  Geophys. Res., 115, doi:10.1029/2010JA015400

\bibitem[{Goncharenko {et~al.}(2013)Goncharenko, Hsu, Brum, Zhang, \&
  Fentzke}]{Goncharenko_etal13}
Goncharenko, L.~P., Hsu, V.~W., Brum, C. G.~M., Zhang, S.-R., \& Fentzke, J.~T.
  2013, J. Geophys. Res. Space Physics, 118, doi:10.1029/2012JA018251

\bibitem[{Heale {et~al.}(2014)Heale, Snively, Hickey, \& Ali}]{Heale_etal14}
Heale, C.~J., Snively, J.~B., Hickey, M.~P., \& Ali, C.~J. 2014, J. Geophys.
  Res. Space Physics, 119, 3857

\bibitem[{Hertzog {et~al.}(2008)Hertzog, Boccara, Vincent, Vial, \&
  Cocquerez}]{Hertzog_etal08}
Hertzog, A., Boccara, G., Vincent, R.~A., Vial, F., \& Cocquerez, P. 2008, J.
  Atmos. Sci., 65, 3056

\bibitem[{Hickey \& Cole(1988)}]{HickeyCole88}
Hickey, M.~P., \& Cole, K.~D. 1988, J. Atmos. Terr. Phys., 50, 689

\bibitem[{Hickey {et~al.}(2010)Hickey, Walterscheid, \&
  Schubert}]{Hickey_etal10b}
Hickey, M.~P., Walterscheid, R.~L., \& Schubert, G. 2010, J. Geophys. Res.
  Space Physics, 115, doi:10.1029/2009JA014927

\bibitem[{Hickey {et~al.}(2011)Hickey, Walterscheid, \&
  Schubert}]{Hickey_etal11}
---. 2011, J. Geophys. Res., 116, doi:10.1029/2011JA016792

\bibitem[{Hines(1960)}]{Hines60}
Hines, C.~O. 1960, Can. J. Phys., 38, 1441

\bibitem[{Hines(1997)}]{Hines97a}
---. 1997, J. Atmos. Sol.-Terr. Phys., 59, 371

\bibitem[{Holton(1976)}]{Holton76}
Holton, J.~R. 1976, J. Atmos. Sci., 33, 1639

\bibitem[{Holton(1980)}]{Holton80}
---. 1980, Ann. Rev. Earth and Planet. Sci., 8, 169

\bibitem[{Holton(1983)}]{Holton83}
---. 1983, J. Atmos. Sci., 40, 2497

\bibitem[{Holton \& Hakim(2012)}]{HoltonHakim12}
Holton, J.~R., \& Hakim, G.~J. 2012, An Introduction to dynamic meteorology,
  5th edn. (Academic Press)

\bibitem[{Jin {et~al.}(2012)Jin, Miyoshi, Pancheva, Mukhtarov, Fujiwara, \&
  Shinagawa}]{Jin_etal12}
Jin, H., Miyoshi, Y., Pancheva, D., {et~al.} 2012, J. Geophys. Res., 117,
  doi:10.1029/2012JA017650

\bibitem[{Jonah {et~al.}(2014)Jonah, de~Paula, Kherani, Dutra, \&
  Paes}]{Jonah_etal14}
Jonah, O.~F., de~Paula, E.~R., Kherani, E.~A., Dutra, S. L.~G., \& Paes, R.~R.
  2014, J. Geophys. Res. Space Physics, 119, 4973

\bibitem[{Kalisch {et~al.}(2014)Kalisch, Preusse, Ern, Eckermann, \&
  Riese}]{Kalisch_etal14}
Kalisch, S., Preusse, P., Ern, M., Eckermann, S.~D., \& Riese, M. 2014, J.
  Geophys. Res. Atmos., 119, 10,081

\bibitem[{Klostermeyer(1972)}]{Klostermeyer72}
Klostermeyer, J. 1972, Z. Geophysik, 38, 881

\bibitem[{Kurihara {et~al.}(2010)Kurihara, Ogawa, Oyama, Nozawa, Tsutsumi,
  Hall, Tomikawa, \& Fujii}]{Kurihara_etal10}
Kurihara, J., Ogawa, Y., Oyama, S., {et~al.} 2010, Geophys. Res. Lett., 37,
  doi:10.1029/2010GL043643

\bibitem[{Labitzke(1981)}]{Labitzke81}
Labitzke, K. 1981, J. Geophys. Res., 86, 9665

\bibitem[{La\v{s}tovi\v{c}ka(2006)}]{Lastovicka06}
La\v{s}tovi\v{c}ka, J. 2006, J. Atmos. Sol.-Terr. Phys., 68, 479

\bibitem[{Limpasuvan {et~al.}(2012)Limpasuvan, Richter, Orsolini, Stordal, \&
  Kvissel}]{Limpasuvan_etal12}
Limpasuvan, V., Richter, J.~H., Orsolini, Y.~J., Stordal, F., \& Kvissel, O.-K.
  2012, J. Atmos. Sol.-Terr. Phys., 78--79, 84

\bibitem[{Lindzen(1981)}]{Lindzen81}
Lindzen, R.~S. 1981, J. Geophys. Res., 86, 9707

\bibitem[{Liu {et~al.}(2011)Liu, Doornbos, Yamamoto, \& Ram}]{Liu_etal11}
Liu, H., Doornbos, E., Yamamoto, M., \& Ram, S.~T. 2011, Geophys. Res. Lett.,
  38, doi:10.1029/2011GL047898

\bibitem[{Liu \& Roble(2002)}]{LiuRoble02}
Liu, H.-L., \& Roble, R.~G. 2002, J. Geophys. Res., 107, 15

\bibitem[{Manney {et~al.}(2009)Manney, Schwartz, Krüger, Santee, Pawson, Lee,
  Daffer, Fuller, \& Livesey}]{Maney_etal09}
Manney, G.~L., Schwartz, M.~J., Krüger, K., {et~al.} 2009, Geophys. Res.
  Lett., 36, doi:10.1029/2009GL038586

\bibitem[{Manson {et~al.}(2002)Manson, Meek, Koshyk, Franke, Fritts, Riggin,
  Hall, Hocking, MacDougall, Igarashi, \& Vincent}]{Manson_etal02}
Manson, A.~H., Meek, C.~E., Koshyk, J., {et~al.} 2002, J. Atmos. Sol.-Terr.
  Phys., 64, 65

\bibitem[{Matsuno(1970)}]{Matsuno70}
Matsuno, T. 1970, J. Atmos. Sci., 27, 871

\bibitem[{Matsuno(1971)}]{Matsuno71}
---. 1971, J. Atmos. Sci., 28, 1479

\bibitem[{Matsuno(1982)}]{Matsuno82}
---. 1982, J. Meteor. Soc. Japan, 60, 215

\bibitem[{Matsuo {et~al.}(2003)Matsuo, Richmond, \& Hensel}]{Matsuo_etal03}
Matsuo, T., Richmond, A.~D., \& Hensel, K. 2003, J. Geophys. Res., 108(A1),
  doi:10.1029/2002JA009429

\bibitem[{McFarlane(1987)}]{McFarlane87}
McFarlane, N.~A. 1987, J. Atmos. Sci., 44, 1775

\bibitem[{Medvedev \& Klaassen(1995)}]{MedvedevKlaassen95}
Medvedev, A.~S., \& Klaassen, G.~P. 1995, J. Geophys. Res., 100, 25841

\bibitem[{Medvedev \& Klaassen(2000)}]{MedvedevKlaassen00}
---. 2000, J. Atmos. Sol.-Terr. Phys., 62, 1015

\bibitem[{Medvedev \& Klaassen(2001)}]{MedvedevKlaassen01}
---. 2001, Geophys. Res. Lett., 28, 733

\bibitem[{Medvedev \& Klaassen(2003)}]{MedvedevKlaassen03}
---. 2003, J. Geophys. Res., 108, doi:10.1029/2002JD002504

\bibitem[{Medvedev {et~al.}(1998)Medvedev, Klaassen, \&
  Beagley}]{Medvedev_etal98}
Medvedev, A.~S., Klaassen, G.~P., \& Beagley, S.~R. 1998, Geophys. Res. Lett.,
  25, 509

\bibitem[{Medvedev \& Yi\u{g}it(2012)}]{MedvedevYigit12}
Medvedev, A.~S., \& Yi\u{g}it, E. 2012, Geophys. Res. Lett., 39,
  doi:10.1029/2012GL050852

\bibitem[{Medvedev {et~al.}(2013)Medvedev, Yi\u{g}it, Kuroda, \&
  Hartogh}]{Medvedev_etal13}
Medvedev, A.~S., Yi\u{g}it, E., Kuroda, T., \& Hartogh, P. 2013, J. Geophys.
  Res. Planets, 118, 1

\bibitem[{Medvedev {et~al.}(2016)Medvedev, Nakagawa, Mockel, Yiğit, Kuroda,
  Hartogh, Terada, Terada, Seki, Schneider, Jain, Evans, Deighan, McClintock,
  Lo, \& Jakosky}]{Medvedev_etal16}
Medvedev, A.~S., Nakagawa, H., Mockel, C., {et~al.} 2016, Geophys. Res. Lett.,
  43, 3095

\bibitem[{Miyoshi \& Fujiwara(2008)}]{MiyoshiFujiwara08}
Miyoshi, Y., \& Fujiwara, H. 2008, J. Geophys. Res., 113,
  doi:10.1029/2007JD008874

\bibitem[{Miyoshi {et~al.}(2014)Miyoshi, Fujiwara, Jin, \&
  Shinagawa}]{Miyoshi_etal14}
Miyoshi, Y., Fujiwara, H., Jin, H., \& Shinagawa, H. 2014, J. Geophys. Res.
  Space Physics, 119, 5807

\bibitem[{Miyoshi {et~al.}(2015)Miyoshi, Fujiwara, Jin, \&
  Shinagawa}]{Miyoshi_etal15}
---. 2015, Journal of Geophysical Research: Space Physics, 120, 10,897,
  2015JA021894

\bibitem[{Palmer(1981)}]{Palmer81}
Palmer, T.~N. 1981, J. Atmos. Sci., 38, 844

\bibitem[{Pancheva \& Mukhtarov(2011)}]{PanchevaMukhtarov11}
Pancheva, D., \& Mukhtarov, P. 2011, J. Atmos. Sol.-Terr. Phys., 73, 1697

\bibitem[{Pancheva {et~al.}(2009)Pancheva, Mukhtarov, Andonov, Mitchell, \&
  Forbes}]{Pancheva_etal09}
Pancheva, D., Mukhtarov, P., Andonov, B., Mitchell, N.~J., \& Forbes, J.~M.
  2009, J. Atmos. Sol.-Terr. Phys., 71, 61

\bibitem[{Pancheva {et~al.}(2008)Pancheva, Mukhtarov, Mitchell, Merzlyakov,
  Smith, Andonov, Singer, Hocking, Meek, Manson, \& Murayama}]{Pancheva_etal08}
Pancheva, D., Mukhtarov, P., Mitchell, N.~J., {et~al.} 2008, J. Geophys. Res.
  Atmos., 113, doi:10.1029/2007JD009011

\bibitem[{Park {et~al.}(2014)Park, L\"uhr, Lee, Kim, Jee, \& Kim}]{Park_etal14}
Park, J., L\"uhr, H., Lee, C., {et~al.} 2014, J. Geophys. Res. Space Physics,
  119, doi:10.1002/2013JA019705

\bibitem[{Pr\"olss(2011)}]{Prolss11}
Pr\"olss, G.~W. 2011, Surv. Geophys., 32, doi:10.1007/s10712-010-9104-0

\bibitem[{Scherhag(1952)}]{Scherhag52}
Scherhag, R. 1952, Ber. Deut. Wetterdienstes, 6, 51

\bibitem[{Schirber {et~al.}(2014)Schirber, Manzini, \&
  Alexander}]{Schirber_etal14}
Schirber, S., Manzini, E., \& Alexander, M.~J. 2014, J. Adv. Model. Earth
  Systems, 6, 264

\bibitem[{Schoeberl(1978)}]{Schoeberl78}
Schoeberl, M.~R. 1978, Rev. Geophys., 16, 521

\bibitem[{Siskind {et~al.}(2010)Siskind, Eckermann, McCormack, Coy, Hoppel, \&
  Baker}]{Siskind_etal10}
Siskind, D.~E., Eckermann, S.~D., McCormack, J.~P., {et~al.} 2010, J. Geophys.
  Res., 115, doi:10.1029/2010JD014114

\bibitem[{Song {et~al.}(2007)Song, Chun, Garcia, , \& Boville}]{Song_etal07}
Song, I.-S., Chun, H.-Y., Garcia, R.~R., , \& Boville, B.~A. 2007, J. Atmos.
  Sci., 34, 2286

\bibitem[{Vadas \& Liu(2009)}]{VadasLiu09}
Vadas, S., \& Liu, H. 2009, J. Geophys. Res., 114, doi:10.1029/2009JA014108

\bibitem[{Vadas \& Fritts(2005)}]{VadasFritts05}
Vadas, S.~L., \& Fritts, D.~C. 2005, J. Geophys. Res., 110,
  doi:10.1029/2004JD005574

\bibitem[{Vadas {et~al.}(2014)Vadas, Liu, \& Lieberman}]{Vadas_etal14}
Vadas, S.~L., Liu, H.-L., \& Lieberman, R.~S. 2014, J. Geophys. Res. Space
  Physics, 119, 7762

\bibitem[{Walterscheid \& Hickey(2012)}]{WalterscheidHickey12}
Walterscheid, R.~L., \& Hickey, M.~P. 2012, J. Geophys. Res., 117,
  doi:10.1029/2011JA017451

\bibitem[{Weinstock(1976)}]{Weinstock76}
Weinstock, J. 1976, J. Geophys. Res., 81, 633

\bibitem[{Weinstock {et~al.}(2007)Weinstock, Klaassen, \&
  Medvedev}]{Weinstock_etal07}
Weinstock, J., Klaassen, G.~P., \& Medvedev, A.~S. 2007, J. Atmos. Sci., 64,
  1027

\bibitem[{Yamazaki {et~al.}(2015)Yamazaki, Kosch, \& Emmert}]{Yamazaki_etal15b}
Yamazaki, Y., Kosch, M.~J., \& Emmert, J.~T. 2015, Geophys. Res. Lett., 42,
  6180

\bibitem[{Yi\u{g}it(2009)}]{Yigit09}
Yi\u{g}it, E. 2009, PhD thesis, University College London Doctoral Thesis

\bibitem[{Yi\u{g}it {et~al.}(2008)Yi\u{g}it, Aylward, \&
  Medvedev}]{Yigit_etal08}
Yi\u{g}it, E., Aylward, A.~D., \& Medvedev, A.~S. 2008, J. Geophys. Res., 113,
  doi:10.1029/2008JD010135

\bibitem[{Yi\u{g}it \& Medvedev(2009)}]{YigitMedvedev09}
Yi\u{g}it, E., \& Medvedev, A.~S. 2009, Geophys. Res. Lett., 36,
  doi:10.1029/2009GL038507

\bibitem[{Yi\u{g}it \& Medvedev(2010)}]{YigitMedvedev10}
---. 2010, J. Geophys. Res., 115, doi:10.1029/2009JA015106

\bibitem[{Yi\u{g}it \& Medvedev(2012)}]{YigitMedvedev12}
---. 2012, Geophys. Res. Lett., 39, doi:10.1029/2012GL053812

\bibitem[{Yi\u{g}it \& Medvedev(2013)}]{YMbook13}
---. 2013, in Climate and Weather of the Sun-Earth System (CAWSES), ed. F.-J.
  L\"ubken, Springer Atmospheric Sciences (Springer Netherlands), 467--480

\bibitem[{Yi\u{g}it \& Medvedev(2015)}]{YigitMedvedev15}
---. 2015, Adv. Space Res., 55, 983

\bibitem[{Yi\u{g}it {et~al.}(2009)Yi\u{g}it, Medvedev, Aylward, Hartogh, \&
  Harris}]{Yigit_etal09}
Yi\u{g}it, E., Medvedev, A.~S., Aylward, A.~D., Hartogh, P., \& Harris, M.~J.
  2009, J. Geophys. Res., 114, doi:10.1029/2008JD011132

\bibitem[{Yi\u{g}it {et~al.}(2012{\natexlab{a}})Yi\u{g}it, Medvedev, Aylward,
  Ridley, Harris, Moldwin, \& Hartogh}]{Yigit_etal12b}
Yi\u{g}it, E., Medvedev, A.~S., Aylward, A.~D., {et~al.} 2012{\natexlab{a}}, J.
  Atmos. Sol.-Terr. Phys., 90--91, 104

\bibitem[{Yi\u{g}it {et~al.}(2014)Yi\u{g}it, Medvedev, England, \&
  Immel}]{Yigit_etal14}
Yi\u{g}it, E., Medvedev, A.~S., England, S.~L., \& Immel, T.~J. 2014, J.
  Geophys. Res. Space Physics, 119, doi:10.1002/2013JA019283

\bibitem[{Yi\u{g}it {et~al.}(2015)Yi\u{g}it, Medvedev, \&
  Hartogh}]{Yigit_etal15a}
Yi\u{g}it, E., Medvedev, A.~S., \& Hartogh, P. 2015, Geophys. Res. Lett., 42,
  doi:10.1002/2015GL064275

\bibitem[{Yi\u{g}it \& Ridley(2011{\natexlab{a}})}]{YigitRidley11a}
Yi\u{g}it, E., \& Ridley, A.~J. 2011{\natexlab{a}}, J. Atmos. Sol.-Terr. Phys.,
  73, 592

\bibitem[{Yi\u{g}it \& Ridley(2011{\natexlab{b}})}]{YigitRidley11b}
---. 2011{\natexlab{b}}, J. Geophys. Res., 116, doi:10.1029/2011JA016714

\bibitem[{Yi\u{g}it {et~al.}(2012{\natexlab{b}})Yi\u{g}it, Ridley, \&
  Moldwin}]{Yigit_etal12a}
Yi\u{g}it, E., Ridley, A.~J., \& Moldwin, M.~B. 2012{\natexlab{b}}, J. Geophys.
  Res., 117, doi:10.1029/2012JA017596

\bibitem[{Yuan {et~al.}(2012)Yuan, Thurairajah, She, Chandran, Collins, \&
  Krueger}]{Yuan_etal12}
Yuan, T., Thurairajah, B., She, C.-Y., {et~al.} 2012, J. Geophys. Res., 117,
  doi:10.1029/2011JD017142

\end{thebibliography}
\end{document}